\documentclass[12pt]{article}
\usepackage[utf8]{inputenc}
\usepackage[margin=1in]{geometry}
\usepackage{amsmath}
\usepackage{wrapfig}
\usepackage{graphicx}
\usepackage{enumitem}
\usepackage[normalem]{ulem} 
\usepackage[style=phys]{biblatex}
\usepackage[table,usenames,dvipsnames]{xcolor}
\usepackage[font=scriptsize, labelfont=bf]{caption, subcaption}
\usepackage{tikz}
\usetikzlibrary{decorations.pathreplacing,calligraphy}
\usepackage{siunitx}
\usepackage{wasysym}
\usepackage{comment}
\usepackage{lineno}
\usepackage{authblk}

\title{Coriolis force deflects wind plant wakes left in the Northern Hemisphere}

\author[1]{Natalie Violetta Frank}
\author[2]{Maria Eletta Negretti*}
\author[2]{Joel Sommeria*}
\author[2]{Mart\'{i}n Obligado*}
\author[1]{Ra\'{u}l Bayo\'{a}n Cal*}

\affil[1]{Department of Mechanical and Materials Engineering, Portland State University, Portland, OR 97201, USA}
\affil[2]{Universite Grenoble Alpes, CNRS, Grenoble-INP, LEGI, 38000 Grenoble, France}

\date{}

\begin{document}

\maketitle

\section*{Abstract:} The physical growth of wind energy over several decades, caused a re-investigation mesoscale effects on wind plant dynamics. At a large scale operation, understanding the interactions between atmospheric phenomena and wind plant wakes is crucial to improving models. Within a wind plant, wakes are generated by individual turbines and the summation of individual wakes is the global plant wake. In both cases, the dynamics are sensitive to atmospheric conditions; one of interest to wind and atmospheric scientists is the Coriolis force. Current research noted spatial and temporal influences on the global plant wakes caused by the Coriolis force. Due to the presence of wind veer in the atmospheric boundary layer, field research is notoriously difficult. As a result, investigations have been lead by numerical simulations. This proved helpful for initial queries, established the significance of the Coriolis force as a non trivial parameter of wind plant dynamics. This works presents a novel experimental study of the impact of the Coriolis force on the dynamics of a scaled wind plant. The experiments show that for a single turbine the wake deflection is insignificant. Additionally, the results show that the global plant wake deflects in the anti-clockwise direction in the Northern hemisphere. The outcomes of this work yield unique experimental methods for research in wind energy, and provide a new basis for the fundamental dynamics of a wind plant under influence of the Coriolis force for future wind plant models causing influence on the global expansion of wind energy. 

\section*{Summary:} In recent years, wind plants and wind turbines have grown so large, that the inclusion of the Coriolis force has become important to wind energy research. However, due to the complications of studying isolated impacts of the Coriolis force in the field, research has been limited to numerical simulations. Whilst it has been ascertained that the Coriolis force has an impact on the wind plant wake, the induced sense of deflection (clockwise or anti-clockwise) remains largely debated. Here, we show experimental results on the influence of isolated Coriolis forces on the dynamics of a wind plant. An anti-clockwise deflection was observed for a wind plant in the Northern hemisphere. 

\section*{Introduction:} 

The growth in renewable energy over recent years has been rapid, with wind energy leading as one of the fastest growing sources of clean energy \supercite{saidur2010review}. This was the result of collective effort to mitigate the effects of global climate change, a consequence of burning fossil fuels \supercite{allen2018chapter}. The advancement of wind turbine technology is driven by the concept that the size of a wind turbine can have a significant impact on the power production. The available power in the wind is influenced by the blade swept area of a wind turbine \supercite{tong2010wind}, referring to the circular area that is swept by the rotating blades of the turbine. This has lead to massive scales of wind turbines and wind plants that can be seen today, with rotors reaching up to $200\text{m}$ in diameter \supercite{irena2019future}, and wind plants spreading over lengths of tens of kilometers.  

Literature has also cited that proper utilization of wind energy relies on adequate characterization of the wind resource \supercite{feng2015modelling}. Combining the recent physical growth of wind energy, and the importance of characterizing the wind resource, questions arose around how these wind plants interact with atmospheric flows from a mesoscale perspective. \textit{In situ} evidence of far-field wakes of a wind plant confirm that these systems have dynamics on a mesoscale within the atmospheric boundary layer (ABL) \supercite{platis2018first}, with wakes extending $45\text{km}$ beyond the offshore wind plant. Wind plants have also been seen to modify the outflow boundary of a thunderstorm slowing the outflow by $20\text{km}\,\text{h}^{-1}$ \cite{tomaszewski2021observations}. Both findings contribute to the argument that wind plants have gotten big enough to see interactions and influences on the atmospheric mesoscale flows. These flows are dominated by both the Coriolis force and the vertical (temperature) stratification. Their importance relative to inertial forces is measured by the non-dimensional Rossby ($Ro$) and Froude ($Fr$) numbers, respectively \supercite{monin1970atmospheric}.  The Coriolis force has a known impact on mesoscale bodies, resulting in flow redirection and pressure drop asymmetry across mountains  \supercite{hunt2001coriolis}. Surface roughness changes also see influence due to the presence of the Coriolis force \supercite{hunt2004coriolis}, and this is seen in flows along coasts \supercite{orr2005coriolis}. Now knowing the importance of considering the Coriolis force on large bodies in the atmospheric boundary layer, wind energy research begins to look at the relevance to wind plant dynamics, as plants can be adapted as a surface roughness change \supercite{crespo1999survey} and turbines and wind plant footprints can now be considered large enough bodies in mesoscale flow \supercite{platis2018first}.

The Coriolis force acts perpendicular to the direction of motion of a moving body, deflecting the trajectory right in the northern hemisphere and left in the southern hemisphere \supercite{persson1998we}. The Coriolis force increases with latitude, making the effect felt strongest at the poles of Earth \supercite{persson1998we}. Although wind plants are being installed all around the world at varying latitudes, there is a high density of wind plants installed around the North Sea \supercite{ewea2009} above mid-latitude conditions which experiences a strong Coriolis force. A preliminary wind resource assessment project, Global Wind Atlas, has developed wind power density and mean wind speed maps; both provide insight into available wind resource and reveal regions of high wind power potential near the poles of the globe \supercite{atlas2021global}. Understanding the influence of the Coriolis force in large wind power potential of northern regions and at high latitudes is important to improving the performance of wind plants. 

For wind plants placed at high latitudes (i.e. the North Sea), and large wind power potential located in northern regions, understanding the Coriolis force's influence in these regions becomes important to improving the performance of wind plants. 

There is a range of consideration when it comes to the Coriolis force in wind plant modelling, and those that approach wind plant modelling from an atmospheric perspective include it i.e. mesoscale models of wind plants \supercite{fitch2012local} and some large eddy simulation (LES) models \supercite{calaf2010large}. As a result, the nuances of the influence of the Coriolis force on wind turbine wake behavior are being further examined. A numerical investigation discovered important insights when considering the Coriolis term, and suggest that it should not be neglected when interactions between plants are considered \supercite{van2015predicting}. Another numerical study demonstrated that power production and wake behavior are impacted by the Coriolis force \supercite{porte2013numerical}. One aspect of wake behavior that is important to wind plant control and revealed to be nontrivial is wake deflection direction. Several numerical studies have investigated this and found the deflection in the clockwise direction can be found in cases of a neutral atmospheric boundary layer (NABL) \supercite{van2015predicting,gadde2019effect}, neutrally stratified atmospheric boundary layers with wind plants modelled as actuator disks \supercite{van2017coriolis}, and in forced boundary layer techniques when wind veer, a Coriolis correction term, and a combination of both are considered in the model \supercite{eriksson2019impact}. On the contrary, other numerical simulations have found that anti-clockwise deflection happens when you consider geostrophic wind direction \supercite{howland2020coriolis}, stably stratified conditions \supercite{dorenkamper2015impact}, conventionally neutral boundary layer \supercite{allaerts2017boundary}, and when the wind plant is modelled as a surface roughness change \supercite{mitraszewski2012wall} in a neutrally stratified atmospheric boundary layer \supercite{van2017coriolis}. All of this to note that the direction of wake deflection due to the Coriolis force is nontrivial and depends on many components. To build on the knowledge of this behavior, the research discussed today is an experimental approach to isolating the effects of the Coriolis force on the wakes within and beyond a wind plant. The experiments discussed neglect stratification of the flow in order to consider sole influence of the Coriolis force. Note, stratification combined with the Coriolis force is important to atmospheric boundary layer characterization, but added complexity to the experimental set up. 

\begin{figure}[h]
    \centering
    \begin{subfigure}{0.49\textwidth}
        \includegraphics[width =\textwidth]{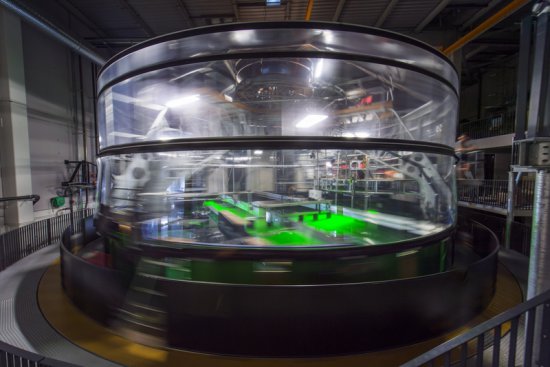}
        \caption{}
        \label{platform}
    \end{subfigure}
    \begin{subfigure}{0.49\textwidth}
        \includegraphics[width =\textwidth]{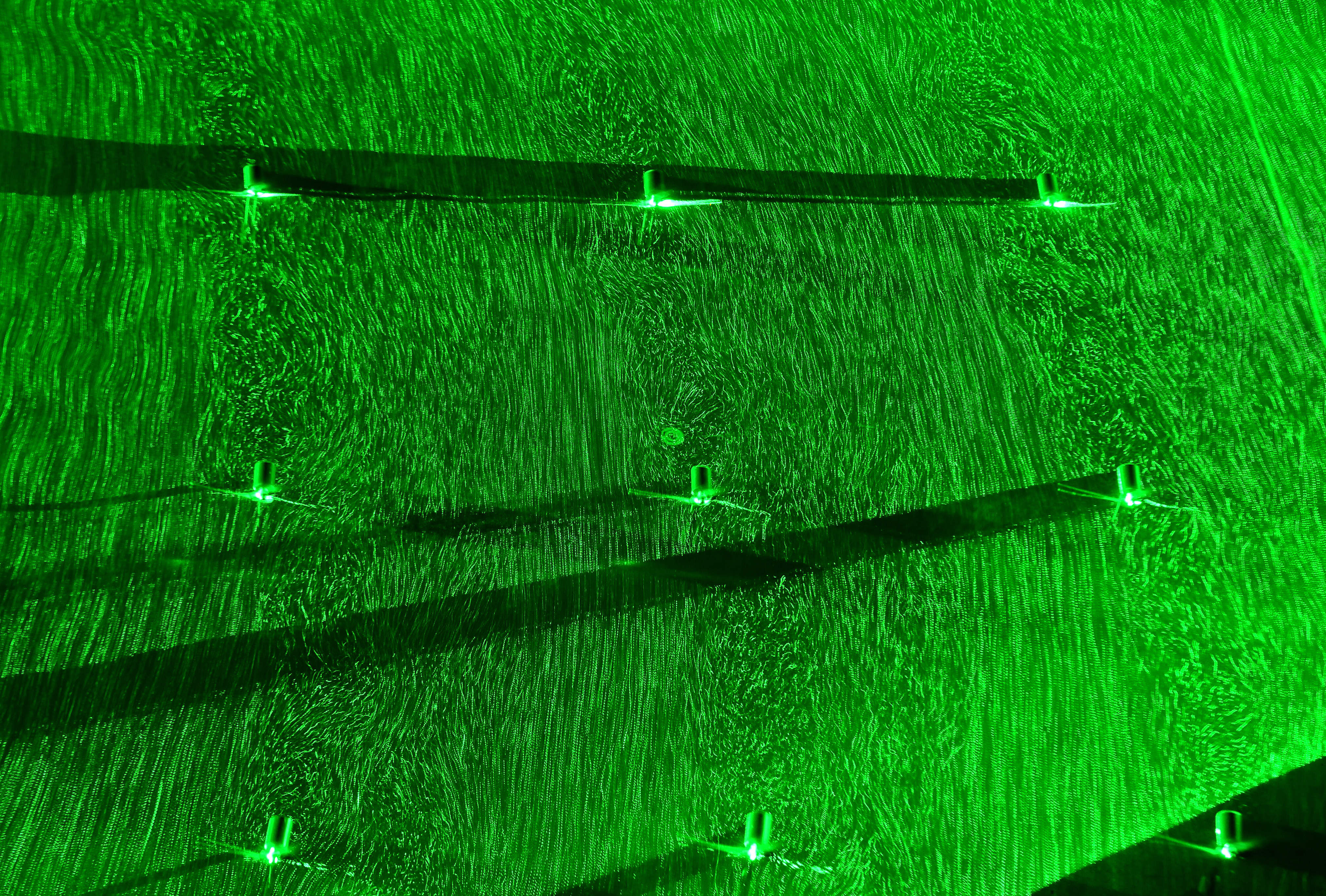}
        \caption{}
        \label{farmim}
    \end{subfigure}
    \caption{a) Rotating Coriolis platform at Universit\'{e} Grenoble Alpes. Long exposure image of the Coriolis platform while it was rotating, the platform spans $13 \text{m}$ in diameter. b) Zoomed in top view photograph of the wind plant experiment. Nine out of twelve turbines are seen. Rotational flow and wakes of the turbines are illuminated by the particles and lasers used in the set up. }
    \label{fancy pictures}
\end{figure}

Mesoscale atmospheric flows are characterized by large Reynolds numbers $Re = \mathcal{O}(10^8)$ and low Rossby numbers $Ro < 1$. Achieving these scales in a laboratory setting is notoriously difficult: usual laboratory experiments cannot reach the typical dynamic regimes characteristic of atmospheric mesoscale flows. The Coriolis rotating platform at LEGI (Grenoble), unique in the world by its size of $13\text{m}$ in diameter, reproduces these flows in dynamic similarity for the main driving forces of buoyancy and rotation of the Earth, with weak influence of viscosity and the centrifugal force. A time-lapsed photo of the Coriolis platform at LEGI is shown in figure \ref{fancy pictures}a. A top view close-up of the scaled wind plant that was used in the experiments is seen in figure \ref{fancy pictures}b. 

The work presented is a novel approach to experimental research, considering a scaled wind plant in a large scale testing facility allowing $Re = 2,000,000$ and $Ro = 0.3$ for a scaled wind plant (based on the length of the plant), as well as $Re = 75,000$ and $Ro = 8.13$ for the case of a single turbine (based on the rotor diameter). From the Rossby numbers of the experiments, preliminary assessment states that the scale of a single turbine is too small to see an impact due to the Coriolis force, whereas the wind plant case will see more influence due to the Coriolis force. The study presented opens new possibilities for future research to quantify the impact of the Coriolis force within and following a wind plant. Ultimately, it contributes to improved wind plant design and power production. 
 
\section*{Results:} The experiments have been conducted using the global set-up sketched in figure \ref{layout}. The wind plant consisted of $3 \times 4$ wind turbines as shown in figure \ref{layout}, in three rows (I, II, III) and four columns (a, b, c, d). The wind turbines used for this experiment were a model based on Odemark and Franson\supercite{odemark2013stability}. Previous work has seen that the direction of blade rotation influences wake behavior in the presence of wind veer in the Northern hemisphere\supercite{englberger2020does}. This experimental approach selected a clockwise blade rotation. The installed wind plant is in a grid Cartesian layout. The rotational flow over the experiment and the results shown are in a polar coordinate perspective. In order to ensure that the turbines in the wind plant are not seeing misaligned flow (and thus results are not due to turbine orientation or geometry), each turbine is rotated slightly to face the respective rotational inflow. The circular geometry of the Coriolis platform produced a uniform flow over the wind plant by anti-clockwise rotation, corresponding to the Northern hemisphere, and maintained a stationary inflow velocity of $u_\theta = 50\text{cm}\,\text{s}^{-1}$. The experimental set-up could be equated to any latitude on earth, as long as the Rossby number is matched. The wind plant configuration could be comparable to the latitude at the poles ($f = 1.45\times10^{-4}$), under conditions of $5\text{m}\,\text{s}^{-1}$ wind speed over a wind plant length of $~38\text{km}$. Further scaling explanation is provided in the experimental design. $2,500$ images were taken using Particle Image Velocitmetry methods at a distance from the bottom of $h = 13.5\text{cm}$, the hub height of the scaled wind turbines. Further details are given in the Methods section. 

\noindent\textbf{Mean velocities:} In figure \ref{velcontour} rotational flow moves from left to right over the displayed wind farm, and is broken down into the cylindrical components of velocity (tangential and radial) in figure \ref{velcontour}a and \ref{velcontour}b respectively. Figure \ref{velcontour}a displays the time averaged normalized tangential velocity, $u_\theta/u_{\theta\infty}$ in a polar coordinate system ($r,\theta$) of the wind plant; the inset in figure \ref{velcontour}a is the normalized tangential velocity of a single turbine. Velocity contours are normalized by the free stream case (pure rotational flow with no turbines present), and is denoted as $u_\theta/u_{\theta\infty}$. Literature agrees that the Coriolis force does not influence the wake of a single turbine, the inset shows that there is no significant impact on the wake dynamics of a single turbine. This is concluded by the horizontal line overlaid on the wake to show that there is no change in radial position of the wake, therefore no deflection from the base rotational flow. 

In the main plot of figure \ref{velcontour}a, there are visibly strong wakes following each turbine in the array, and fully merged row wakes within the plant. The wakes behind the first column of turbines (column a) are not changing in radial position, meaning the wakes of these turbines do not deflect with respect to the base rotational flow. The second column of turbines appears to have very little deflection from the initial radial position, and the following columns of the array show strong deflections from the initial turbine radial positions and thus a deflection from the background rotational flow. Deflection seen in this wind farm is not attributed to misaligned flow. In this set-up, turbines are oriented to face perpendicular to the respective rotational inflow. The wakes in these columns of turbines (columns b-d), and the global wake deflection of the normalized $u_\theta$ velocity is in the anti-clockwise direction. 

To explore this deflection further, the time averaged normalized radial velocity is plotted in figure \ref{velcontour}b with an inset of the single turbine case. This velocity contour is normalized by the free stream case (pure rotational flow with no turbines present) and is denoted as $u_r/u_{r\infty}$. A positive value indicates that the radial component of velocity is in the downward direction. This is consistent with the concept of the direction the Coriolis force in the northern hemisphere; pushing moving bodies to the right. In this experiment rotational flow moves from left to right over the scaled wind plant, and thus a positive radial velocity is in the downward direction. In the single turbine case, the normalized velocity in the wake is slightly negative, indicating a slight reversal of the radial velocity. However, the magnitude of the wake is very small, approximately $\leq1$, while the surrounding flow maintains a value of approximately $1$ on either side of the wake. This explains why there is no deflection in the single turbine case, the radial velocity inside and outside the wake is in relative balance with one another and therefore causes no deflection in the presence of the Coriolis force.  

Referring to the main wind plant contour of figure \ref{velcontour}b a similar situation to the single turbine is seen in the first column of the array, and a merged wake is seen beginning at the last column of turbines in the array. The radial velocity immediately behind the first column of turbines are of equal magnitude to the surrounding flow. Starting at the second column a slight re-direction begins and increases in magnitude with each turbine column that follows. The magnitude of the normalized radial velocity in the global wake is $>2$ while the surrounding flow is $\leq 1$. This imbalance in radial velocity is causing re-direction of the wakes in the back of the wind plant, and contributing to the anti-clockwise deflection of the global wind plant wakes. 

Wake center-line tracking was used to visualize the deflection of turbine wakes within the plant. To track the center line of the wake single Gaussian curves were fit to the velocity data. Figure \ref{center_profiles}a is a plot of the difference between the wake center and the initial turbine position in column a as a function of a downstream position, $x$, in the plant. Turbines are located at $0D$, $3D$, $6D$ and $9D$. Behind the first column of turbines in the wind plant, $0D$, there is no difference in wake center from the initial radial position of the turbine, translating to no wake deflection behind this turbine. This is consistent with what was observed in the contour plots of figure \ref{velcontour}a. From $3D$ and beyond, there is a difference between the wake center and the initial turbine position, deflection of the wakes are in the anti-clockwise direction. The deflections of the wakes are quantified and seen clearly in this figure. This information also implies that the wind resource will be different for each turbine in the plant. Investigating this characteristic involves taking velocity profiles behind each turbine in the plant. 

Figure \ref{center_profiles} shows the wake profiles behind the first row of the wind plant. The profiles reveal that there are varying degrees of wake recovery behind each turbine, as well as different profile shapes. This says that although there are many factors that influence the wind resource of each turbine in a wind plant, the Coriolis force should not be neglected. This is important for wind plant modelling and contributes to methods for wind energy production. Both plots in figure \ref{center_profiles} show that the inclusion of the Coriolis force causes changes in the interactions between turbines in a plant, in addition to global wake changes. Dynamic changes on the mesoscale wake and the dynamics of turbine interactions are important scales for wind plant modelling. 

\section*{Discussion}

A strong understanding of the wind resource is needed to maximize plant performance when choosing a location.  This has been the motivation for wind energy scientists for decades. This has lead to extensive studies of the available wind resource around the world. One study done by the Global Wind Atlas shows a region of high potential near the north pole. In addition to this region of high potential, there are currently numerous offshore wind plants in the Northern Sea, located above mid-latitude. The Coriolis force must be considered to gain a full understanding of the wind resource in both cases. 

This atmospheric boundary layer parameter has been an important one for meteorological sciences and the studies of large bodies, such as mountains, and surface roughness changes, such as coasts. With wind plant and turbine sizes growing to massive proportions, along with the trove of research highlighting the importance of the Coriolis force in atmospheric characterization and the influence on wind veer, wind energy science has accepted this parameter as non-trivial and important for wind modelling. Investigations of Coriolis effects on wind turbines and wind plants have been limited to numerical research. Due to the complexity that comes with properly recreating Coriolis effects on a scaled set up, experimental methods have not been performed. The research presented here not only yields quantitative information about the impact of Coriolis effects on wind plant's dynamics, but it opens the door to a new method of research for wind plant dynamics. 

One of the major questions surrounding Coriolis forces and wind plants is the direction of wake deflection. The research presented here gives insight on the matter and it is not a trivial answer. Based on the experiment discussed in this research, the direction of deflection of a global plant wake is anti-clockwise. Investigating deflection behind each column of turbines shows that there is no deflection in the first column of the plant, and that deflection slowly increases with the distance from the entrance of the wind plant. This deflection also causes the wind resource of each turbine in the plant to be slightly different from the one ahead of it.

Understanding the interactions of the turbines in a wind plant and their respective behavior based on the incoming wind is important to the performance and production of the wind plant. By investigating the deflections of each turbine in the plant, we are studying the relative wind resource for the turbines following it. Under the influence of the Coriolis force, the approaching wind of the first turbines in a wind plant is modified from a uniform inflow assumption, being one of the possible influences, in addition to wake effects, on modifying the wind resource of each turbine following the entrance of the farm. Providing information not only on how the Coriolis force will change the wind plant dynamics but also how this new condition changes the characteristics of the wind source for each turbine in a plant is a significant part of this research. The research will present a new experimental method of studying wind plant dynamics under the influence of the Coriolis force and guides new research avenues. This study lays the ground work for future iterations of experiments investigating the influence of stratification and the presence of the Coriolis force and wake recovery. 

\newpage
\section*{Methods:} 
\begin{figure}[h]
    \centering
    \includegraphics[width = \textwidth]{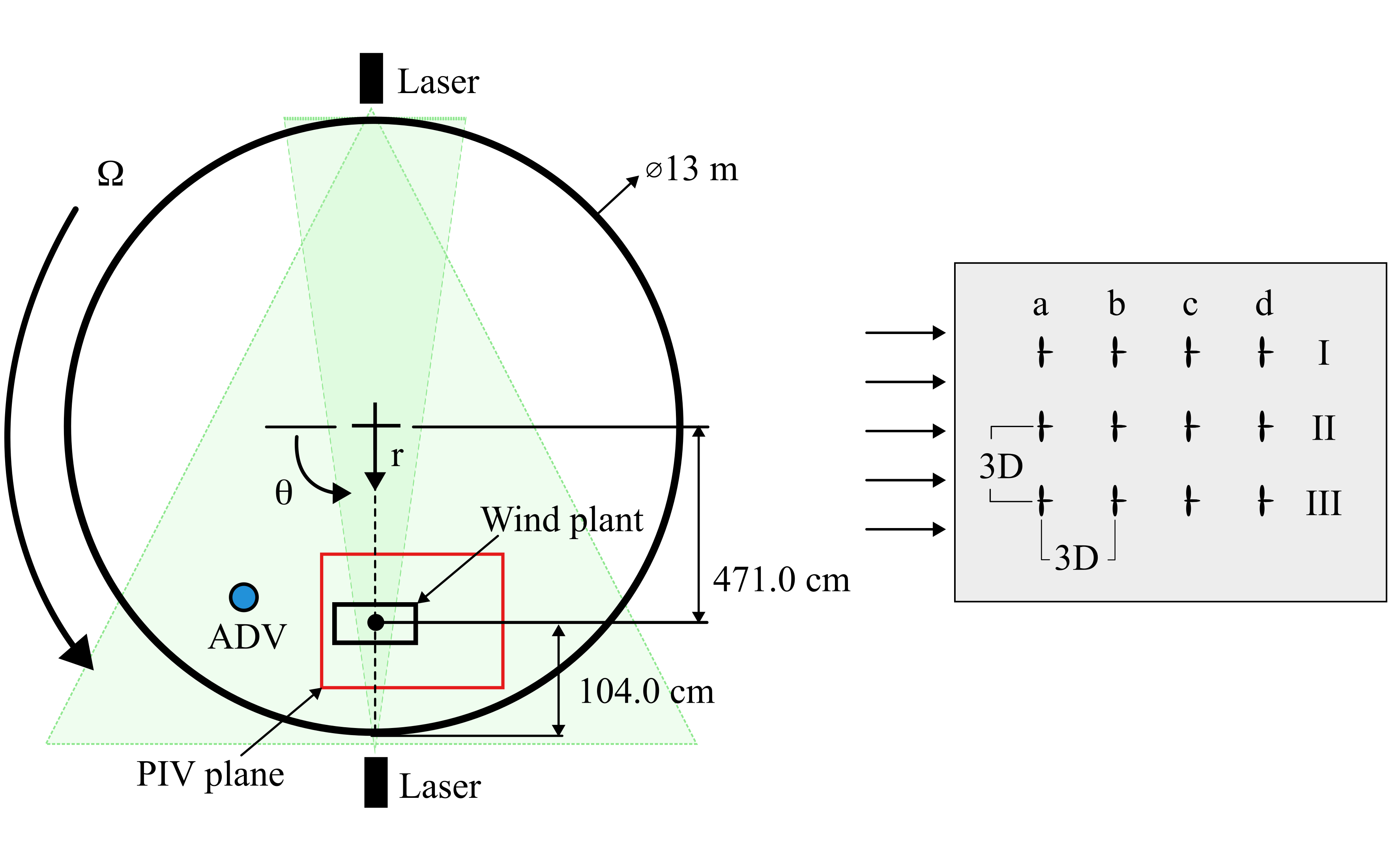}
    \caption{Top view experimental schematic of the Coriolis platform and wind plant layout. The schematic on the left illustrates the experimental set up of the entire Coriolis platform. The direction of rotation of the platform is shown, in the anti-clockwise direction, corresponding to northern hemisphere rotation. Components of the experimental set up are labelled here. The schematic on the right shows the layout of the wind plant. Columns are labelled with letters $a, b, c,$ and $d$. Rows of the wind plant are labelled with roman numerals I, II, III. Further details on placement can be found in the methods section.}
    \label{layout}
\end{figure}

\begin{figure}[h]
\begin{subfigure}{0.49\textwidth}
    \includegraphics[width = \textwidth]{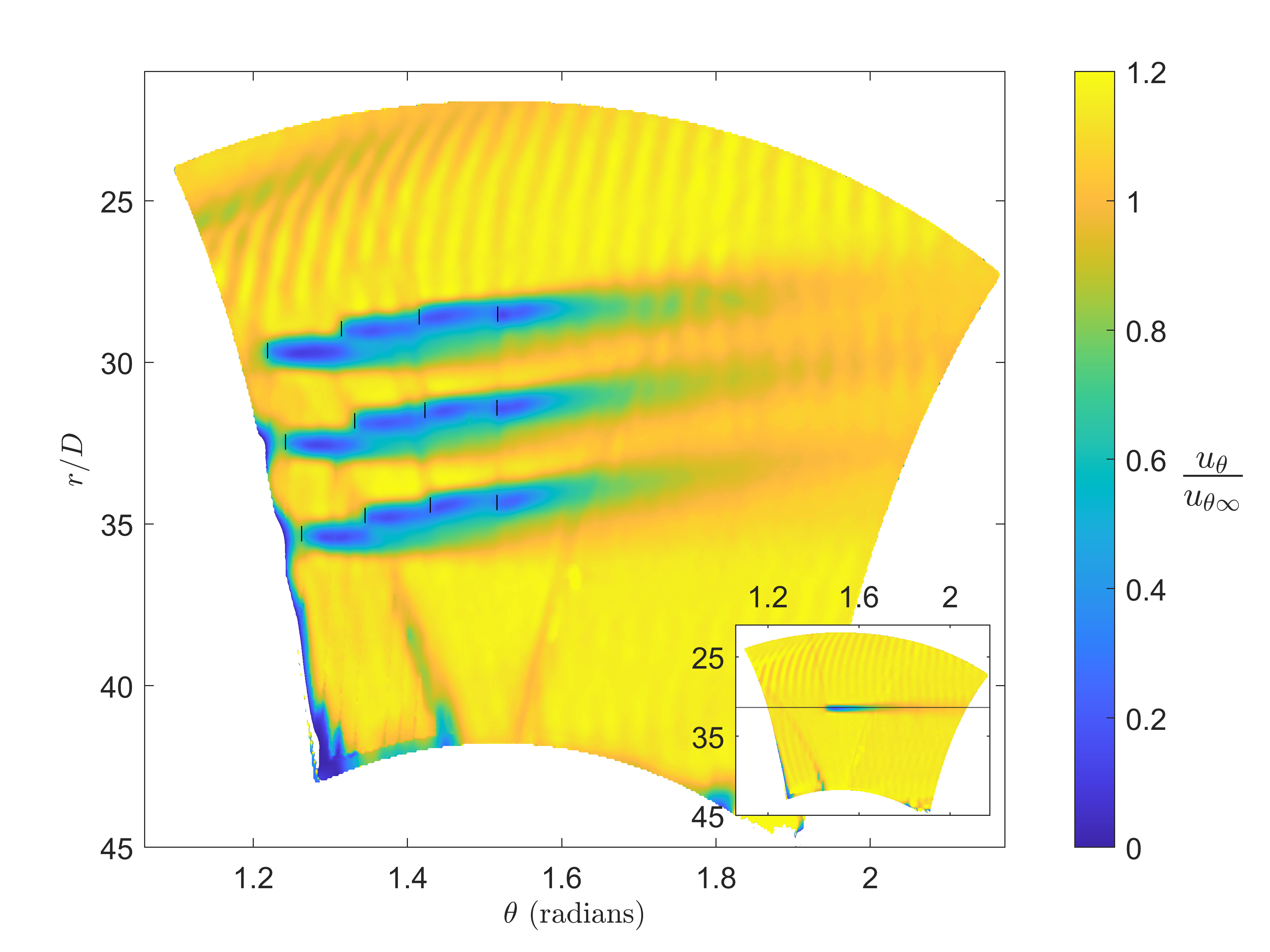}
    \label{utcontour}
    \caption{}
\end{subfigure}
\begin{subfigure}{0.49\textwidth}
    \includegraphics[width=\textwidth]{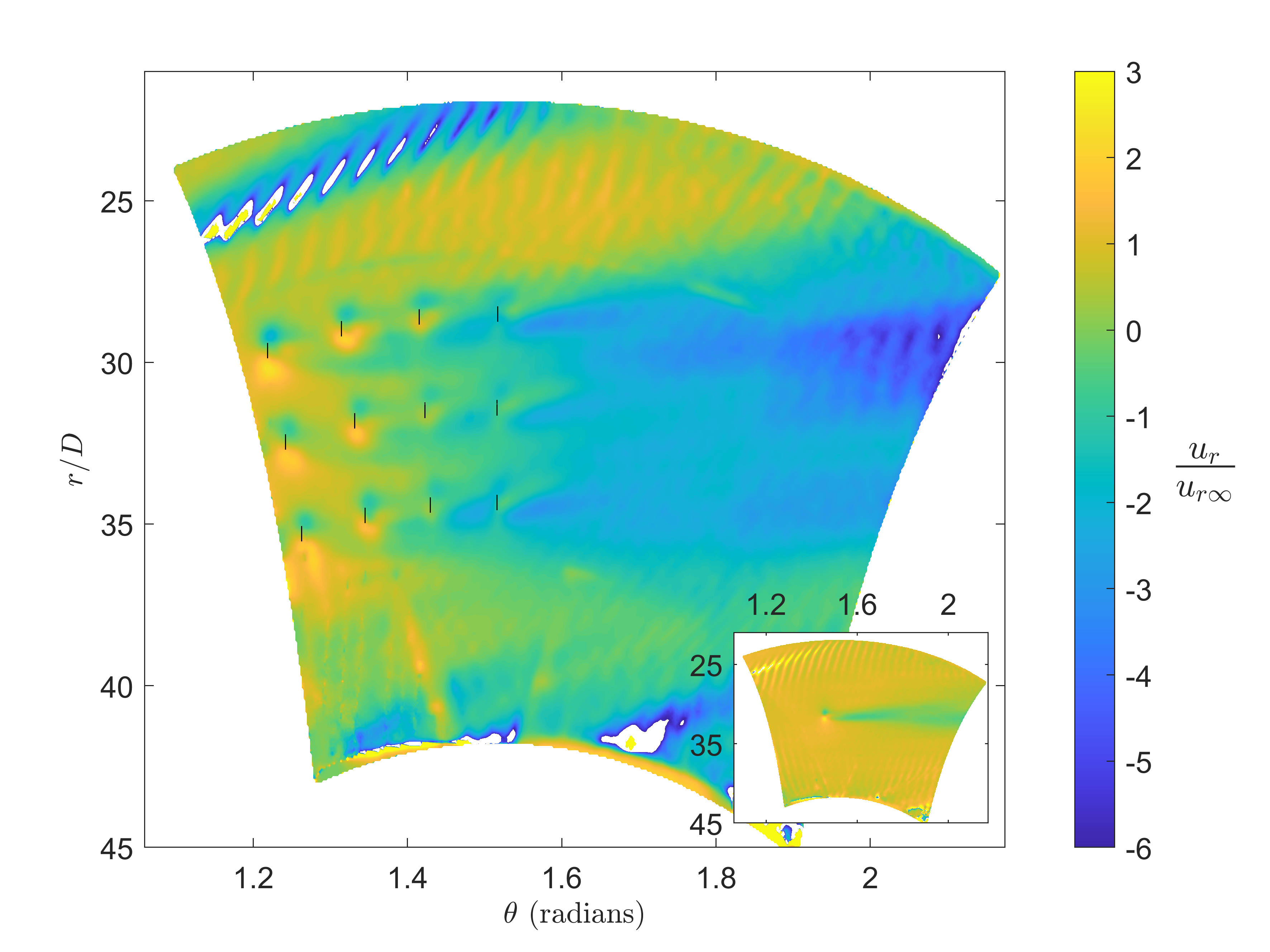}
    \label{urcontour}
    \caption{}
\end{subfigure}
\caption{a) Contour plot of the normalized tangential velocity, ${u_\theta}/{u_{\theta\infty}}$, of the 12 turbine array case in polar coordinates. Vertical dashes indicate turbines. The inset figure is the case of a single turbine subjected to Coriolis force. The horizontal line through the wake of the turbine is to show that there is indeed no wake deflection from the initial radial position of the turbine, and thus no deflection from the background rotational flow. In the main figure, the wakes of the turbines in columns are not horizontal, indicating deflection from the initial radial position of the turbine and thus a deflection from the background rotational flow. These wakes in these figures are directed upwards, signifying an anti-clockwise deflection of the wakes. b) Contour plot of the normalized radial velocity, ${u_r}/{u_{r\infty}}$, of the 12 turbine array case in polar coordinates. Vertical dashes indicate turbines. The inset figure is the case of a single turbine subjected to Coriolis force. In the single turbine contour, the magnitude of the normalized radial velocity in the wake is very small compared to the magnitudes of the wakes in the 12 turbine array case. The magnitude in the single wake is approximately $<1$ and the surrounding flow is approximately $1$. The normalized radial velocity magnitude of the single wake is in relative balance with the magnitude of the surrounding flow, thus creating no deflection in the single turbine case. In the main figure, a similar balance is seen in the first two columns of the array. The last two columns of the array show higher magnitudes in the global wake than the surrounding flow, $>2$ and approximately $1$ respectively. This imbalance in radial velocity magnitude is what causes the large anti-clockwise deflection of the global farm wake. }
\label{velcontour}
\end{figure}

\begin{figure}[h]
\begin{subfigure}{0.47\textwidth}
    \includegraphics[width = \textwidth]{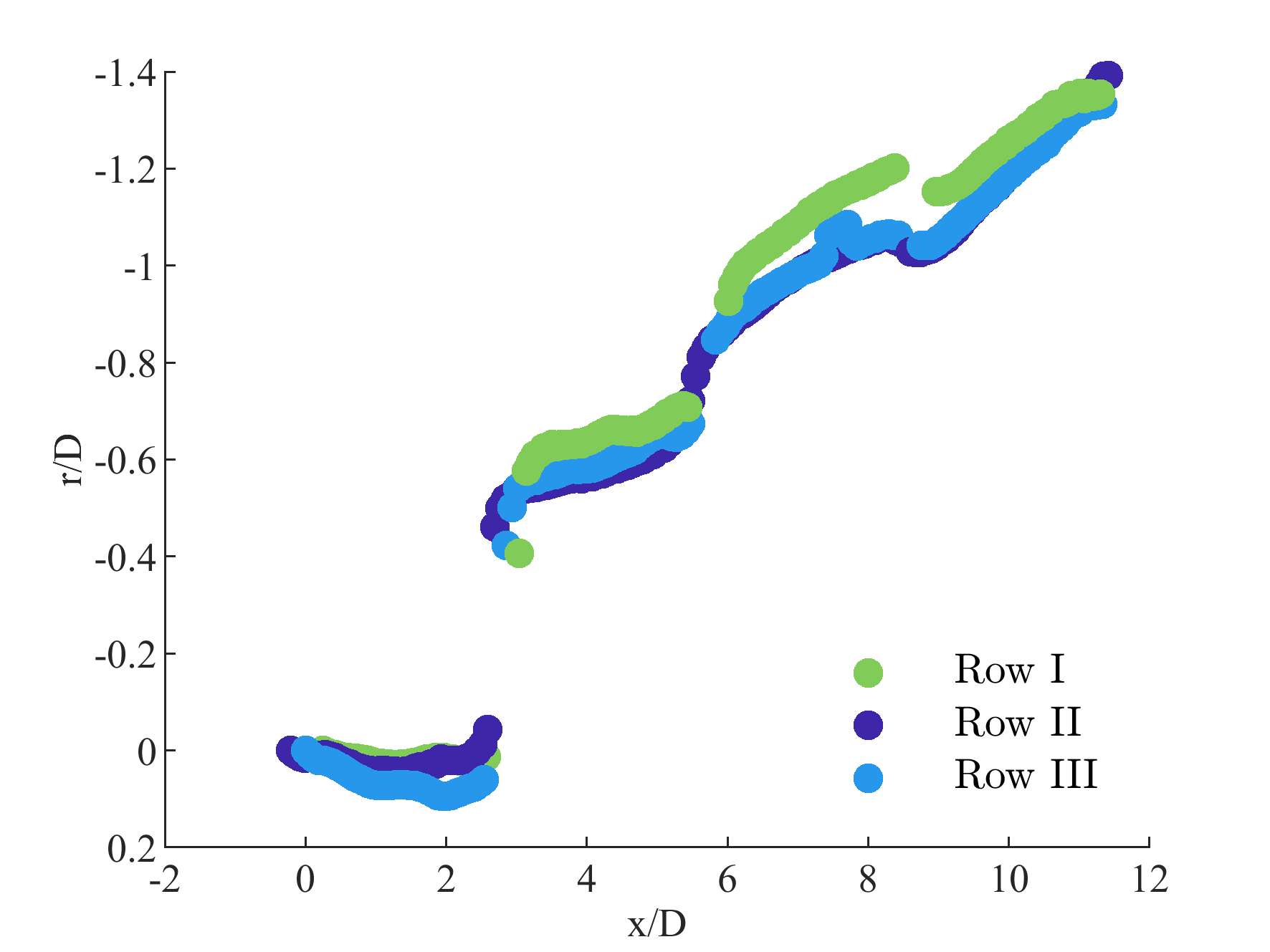}
\end{subfigure}  
\begin{subfigure}{0.47\textwidth}
    \includegraphics[width = \textwidth]{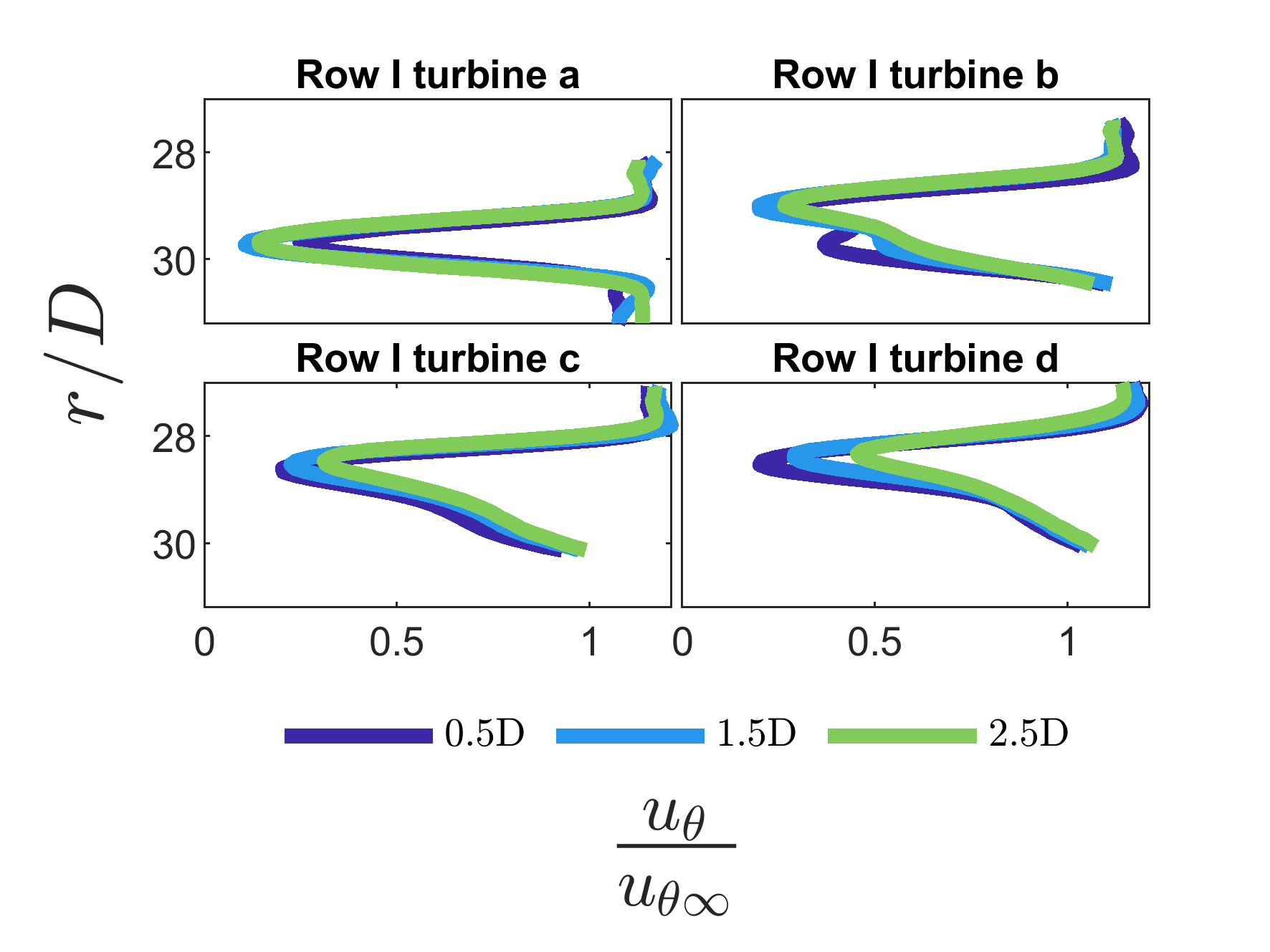}
\end{subfigure}
    \caption{a) Wake center line tracking of each row in the $3D \times 3D$ plant. Turbines are located at $0D, 3D, 6D,$ and $9D$. The figure plots the change in radial distance from the initial turbine position of each row, as a function of downstream position in the wind plant. Values of the x and y axis are normalized by $D$. Behind the first turbines of the wind plant, there is little to no deflection in the wakes. However, all turbines after that have a strong anti-clockwise deflection in the wakes. This deflection is a result of the Coriolis forces in these experiments. b)  Velocity profiles at $0.5D, 1.5D$ and $2.5D$ behind each turbine in row I of the wind plant. There are four panels in this figure, each panel is the velocity profiles of one turbine in the first row of the wind plant. The profiles vary in shape and wake recovery. There is more wake recovery in turbines $b$, $c$, and $d$ than in turbine $a$. b) Velocity profiles at $0.5D, 1.5D$ and $2.5D$ behind each turbine in row I of the wind plant. There are four panels in this figure, each panel is the velocity profiles of one turbine in the first row of the wind plant. The profiles vary in shape and wake recovery. There is more wake recovery in turbines $b$, $c$, and $d$ than in turbine $a$.}
    \label{center_profiles}
\end{figure}

The Coriolis rotating platform at LEGI in Grenoble (France) consists of a circular tank of 13 m diameter and $1\text{m}$ depth. It is the largest rotating tank in the world for geophysical investigations (see www.legi.grenoble-inp.fr/), allowing different stratifications in both temperature and salinity and different rotation speeds. For the present study, the tank was first filled with fresh water to a total depth of $100\text{cm}$. The wind plant consisting of 12 equally spaced turbines and a spacing of $3D$ (with $D=15\text{cm}$ being the rotor diameter) were arranged in 3 rows (I, II, III) and four columns (a, b, c, d) as sketched in figure \ref{layout}. We also considered the case of a single wind turbine placed $470.1\text{cm}$ away from the center of the tank. A three-blade wind turbine model was chosen, with a rotor design based on the model of \cite{odemark2013stability}, scaled to a diameter of $15\text{cm}$ and a hub height of $13.5 \text{cm}$ rotating in the clockwise direction. These dimensions have been chosen in order to assure enough resolution but minimize effects of recirculation of the flow from the free surface, bottom and lateral boundary layers. The rotor blades were 3D printed (Markforged 3D printer) using Onyx filament with fiberglass continuous fibers.

The flow was generated by putting the tank in a differential spin-up solid-body anticlockwise (cyclonic) rotation with an averaged period of $T=30\text{s}\pm5\text{s}$ corresponding to a Coriolis parameter $f=2\Omega=4\pi/T=0.41\text{s}^{-1}$. This produced a nearby stationary and uniform inflow at the plant entrance over $37.5\text{s}$, corresponding to 1.25 rotational days.
Turbines were fixed to the platform by screwing into a large plate secured to the platform by glue. The turbines in the plant, and the single turbine case, were each oriented normal to their respective incoming rotational velocity. The velocity of the inflow was monitored by means of an Acoustic Doppler Velocimetry (ADVP, Vectrino) device, placed at a distance of $469.8 \text{cm}$ from the tank center, and $312.0 \text{cm}$ from the single turbine position (figure \ref{layout}). The averaged ADV incoming flow velocity was $50\text{cm}\,\text{s}^{-1}$ for the entire duration of the experiment. Polyamide particles (Orgasol) with a mean diameter of $60 \mu\text{m}$ and a density of $1.020\text{kg}\,\text{m}^{-3}$ were added to the the water to allow optical velocity measurements using the particle image velocimetry (PIV) technique and obtain two-dimensional instantaneous velocity fields. A first $25 W$ Yag laser operating at a wavelength $\lambda = 532 \text{nm}$ provided a first continuous light source producing a laser sheet at hub height $h=13.5\text{cm}$ from the bottom using a diverging lens of $30^\circ$ close to the wind plant and positioned outside of the rotating tank. The use of a second laser $10 W$ Yag laser operating at the same wavelength with a diverging lens of $30^\circ$ allowed to reduce the shadows produced from the first light source by the wind turbines (cf. figure 1b). 
 
Images of $400 \text{cm} \times 350 \text{cm}$ were taken with a high-resolution JAI Camera (14MPx) synchronized with the laser system, at a frame rate of $66 Hz$. The spatial resolution of $1\text{mm}\,\text{pixel}^{-1}$ was obtained using an optical lens of $14 \text{mm}, F2.8$ on the camera. Velocity fields were computed using a cross-correlation PIV algorithm encoded with the software UVMAT developed at LEGI. For this purpose, an adaptive multi-pass routine was
used, starting with an interrogation window of $35 \times 35$ pixels and a final window size of $20 \times 20$ pixels, with a 70\%
window overlap. Each element of the resulting vector field thus represents an area of roughly $0.5 \times 0.5 \text{cm}$. The
velocity vectors were post-processed using a local filter for erroneous vectors. The maximum experimental error is estimated to be about 3\% in the instantaneous velocity and about 10\% in its spatial derivatives.

\newpage
\printbibliography

\newpage

\newpage

\newpage

\newpage

\newpage
\section*{Acknowledgements}
Special thanks go to Samuel Viboud and Thomas Valran for their continuous support in conducting the experiments at the Coriolis Platform.

\newpage
\section*{Author Information}
\subsection*{Authors and Affiliations}
\textbf{Department of Mechanical and Materials Engineering, Portland State University, Portland, OR 97201, United States of America}\\
Natalie Violetta Frank, Ra\'{u}l Bayo\'{a}n Cal\\

\noindent\textbf{Universit\`{e} Grenoble Alpes, CNRS, LEGI, 38000 Grenoble, France}\\
Maria Eletta Negretti, Joel Sommeria\\

\noindent\textbf{Universit\`{e} Grenoble Alpes, CNRS, Grenoble-INP, LEGI, 38000 Grenoble, France}\\
Martin Obligado

\subsection*{Contributions}
N.V.F., M.O., R.B.C. designed the experimental set up; M.E.N. and J.S. refined the image measurement set up; N.V.F. performed the experiments; J.S. developed methodology to convert raw images of the experiment, to velocity vector field data for analysis; N.V.F., M.E.N., and J.S processed experimental data; N.V.F. analyzed experimental data and developed explanation for anti-clockwise wake deflections; N.V.F. lead the writing of the manuscript with input from all other authors. 

\subsection*{Corresponding Author}
Correspondence to Natalie Violetta Frank.
\newpage
\section*{Competing Interest Declaration}
The authors declare no competing interest. 

\newpage
\section{Supplemental Material: Non-dimensional numbers} 

In the simulation of geophysical phenomena in laboratory tanks, a fundamental point is to achieve both geometrical and dynamical similarity between real-atmospheric and tank phenomena. For the geometric scaling the rotor diameter to hub height ratio of the scaled wind turbine is $1.11$. The scaling
ratio wind turbine is $1 : 780$ corresponding to a real rotor diameter of 117m, and a wind plant area of $1,053\text{m}\times702\text{m}$. 

There are two non-dimensional numbers which are relevant for the dynamical similarity and need to be achieved in the laboratory. These are the Rossby number, $Ro=U/fL$ and the Reynolds number, $Re=UL/\nu$, giving the ratio between the inertial and the Coriolis forces or the viscous forces, respectively. Herein, $U$ is a typical advective velocity scale, $L$ is a characteristic length scale, $f$ is the Coriolis parameter and $\nu$ (1$\cdot10^{-6}\text{m}^2\text{s}^{-1}$ for water and 1.57$\cdot10^{-5}\text{m}^2\text{s}^{-1}$ for air) is the kinematic viscosity. Both numbers, can be defined using two different relevant length scales: the first one is given by the rotor diameter $D$, the second one given by the size of the wind plant $L_p$. The scalings are summarized in table \ref{tab1}.
\begin{table}[h]
    \centering  
\begin{tabular}{|c|c|c|c|}
\hline 
parameter & real model & experiments & ratio\tabularnewline
\hline 
\hline 
$L_p$ &  1.053 km & 1.35 m & 780\tabularnewline
\hline 
$D$ &  117 m & 0.15 m & 780\tabularnewline
\hline 
$h$ &  105.3 m & 0.135 m & 780\tabularnewline
\hline 
$D/h$ &  1.11  & 1.11 & 1\tabularnewline
\hline 
 $U$ & 5 m/s & 0.5 m/s & 10\tabularnewline
 \hline 
$f$ &  $10^{-4}$ $\text{s}^{-1}$& 0.41 $\text{s}^{-1}$ & 2.43$\times10^{-4}$ \tabularnewline
\hline 
 $\mathrm{Ro_D}=U/fD$ & 427 & 8 & 52.5\tabularnewline
\hline 
 $\mathrm{Re_D}=U D/\nu$ & 3.7$\cdot 10^{7}$  & 7.5$\cdot 10^{4}$ & 5$\cdot 10^{2}$\tabularnewline
\hline 
$\mathrm{Ro_{L_p}}=U/fL_p$ & 47 & 0.9 & 52.2\tabularnewline
\hline 
 $\mathrm{Re_{L_p}}=U L_p/\nu$ & 3.3 $\cdot 10^{8}$& 6.75$\cdot 10^{5}$ & 5$\cdot 10^{2}$\tabularnewline
\hline 
\end{tabular} 
\caption{Scaling of relevant parameters to the experimental set up. Values for the real life model and the experimental model are listed. }
    \label{tab1}
\end{table}

\newpage
\section{Supplemental Material: Velocity Profiles}
\begin{figure}[h]
    \centering
    \includegraphics[width = \textwidth]{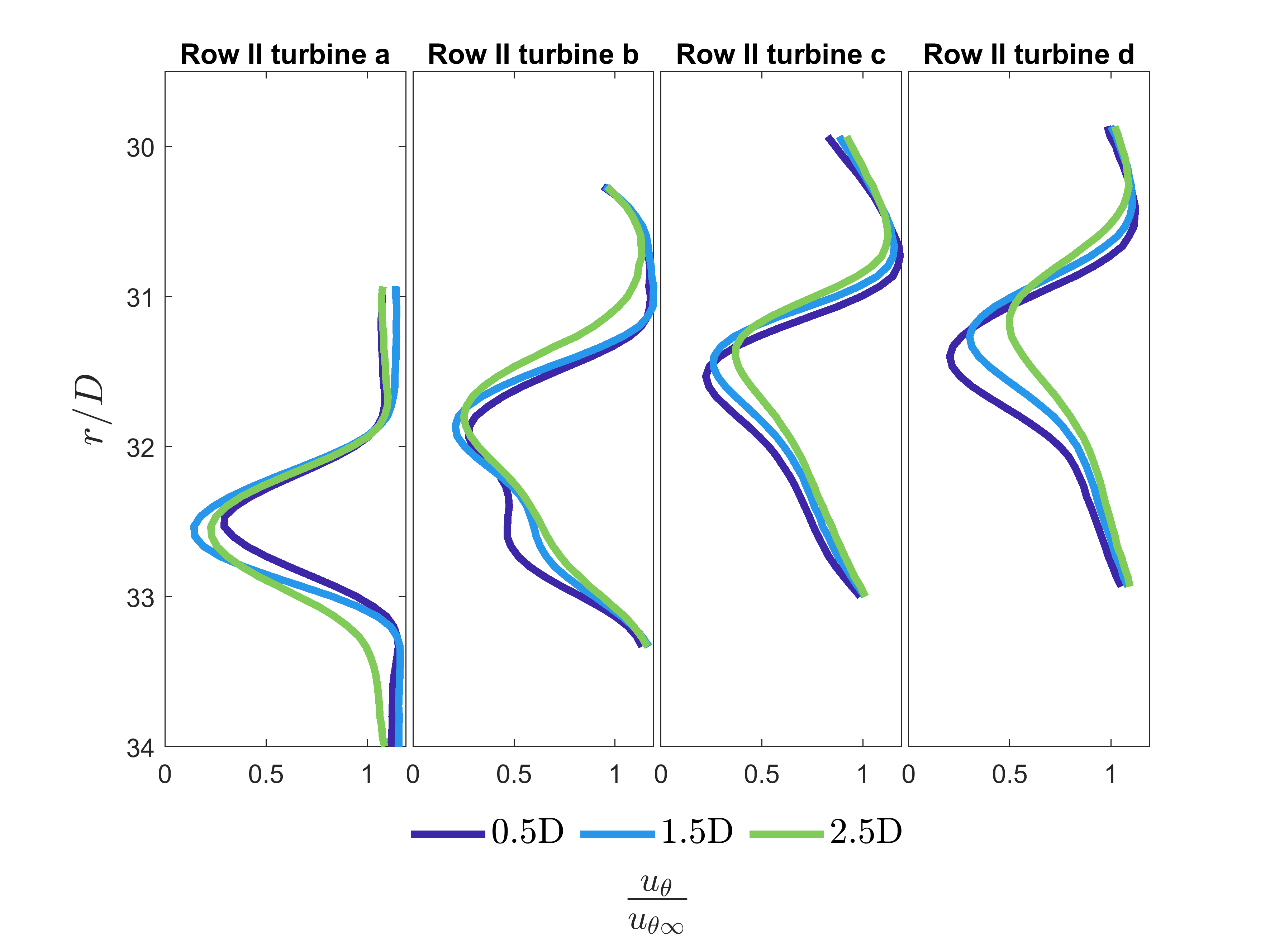}
    \caption{Velocity profiles at $0.5D, 1.5D$ and $2.5D$ behind each turbine in row II of the wind plant. There are four panels in this figure, each panel is labelled at the top to identify the turbines within row II. Velocity profile shapes of row II follow a similar trend to row I. However, there is slightly more wake recovery in turbines $a$ and $b$ in row II than in row I. }
    \label{row2p}
\end{figure}
\begin{figure}[h]
    \centering
    \includegraphics[width = \textwidth]{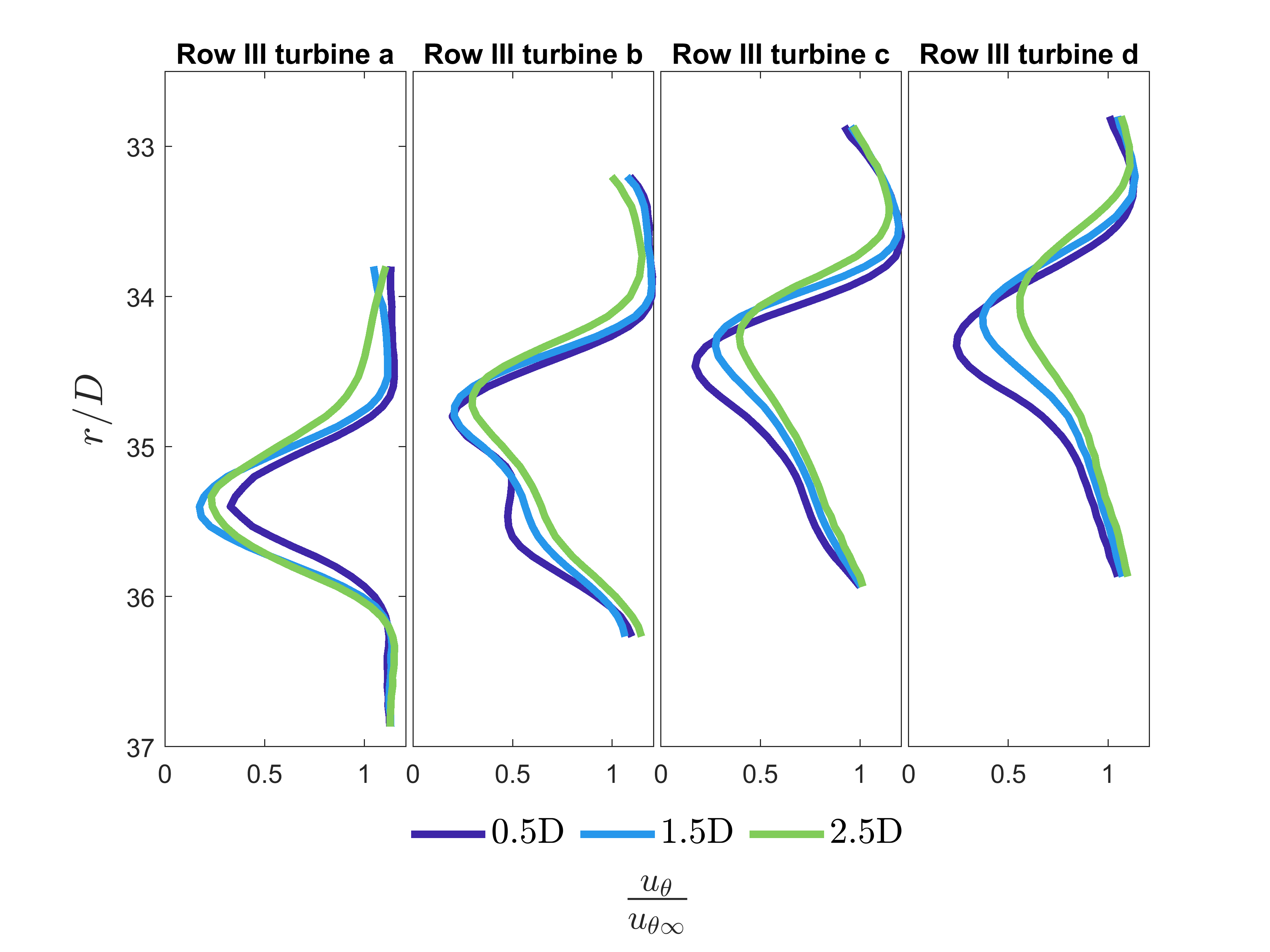}
    \caption{Velocity profiles at $0.5D, 1.5D$ and $2.5D$ behind each turbine in row III of the wind plant. There are four panels in this figure, each panel is labelled at the top to identify the turbines within row III of the plant. Velocity profile shapes follow a very similar trend to the first row of the wind plant. However, there appears to be more wake recovery behind turbines $a$ and $b$ of row III, than behind turbines $a$ and $b$ of row I. }
    \label{row3p}
\end{figure}
\clearpage

\newpage
\section{Supplemental Material: Single Turbine Reference}
\begin{figure}[h]
    \centering
    \includegraphics[width=\textwidth]{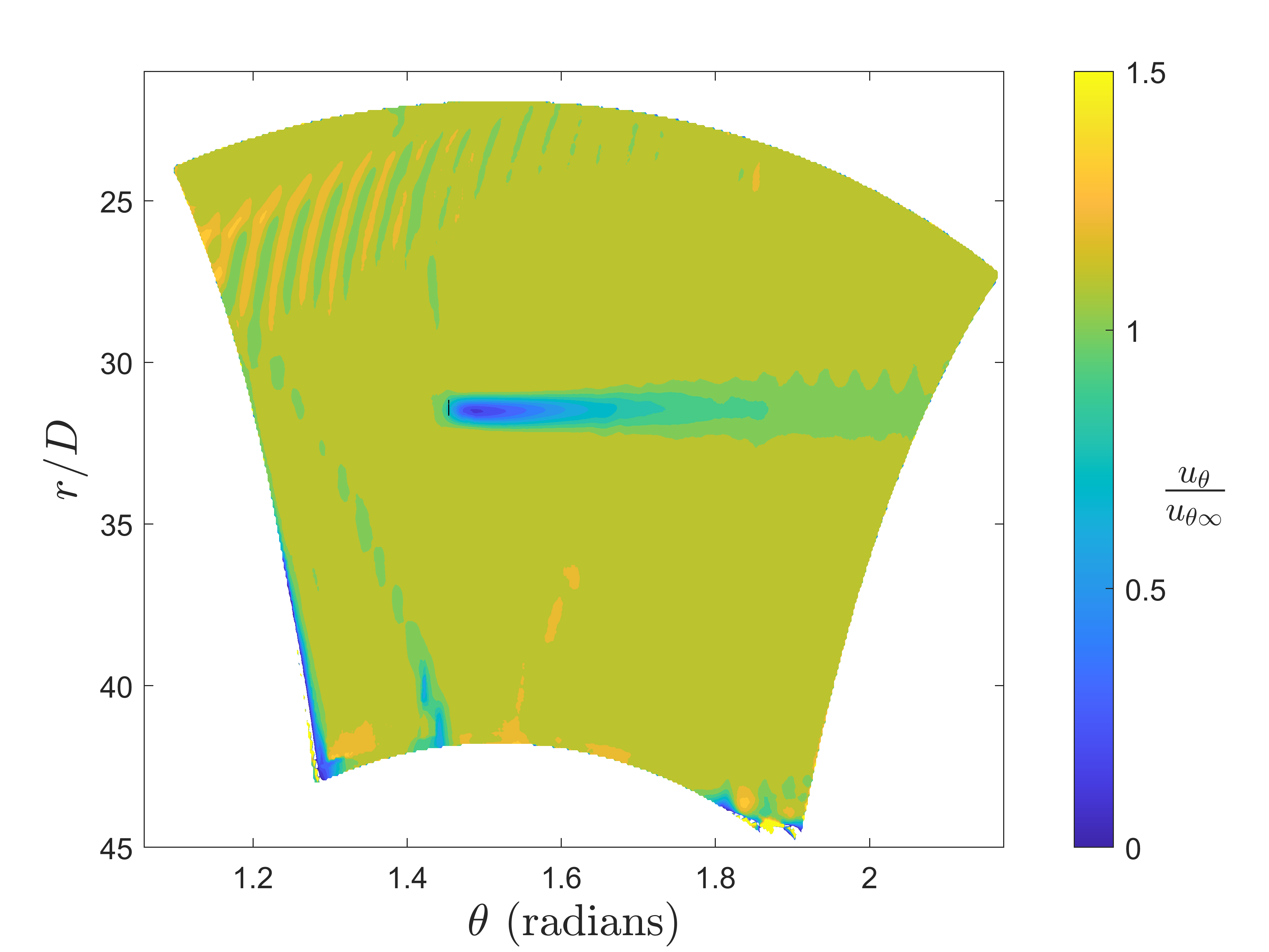}
    \caption{$u_\theta / u_{\theta\infty}$ normalized mean tangential velocity contour plot in polar coordinates of the single turbine case. The origin is located at the center of the tank. $r$ is normalized by the rotor diameter and $u_\theta$ is normalized by the freestream tangential velocity, $u_{\theta\infty}$. From this figure, it can be seen that there is no wake deflection from the radial position behind a single turbine due to Coriolis forces. This is consistent with previous literature that concluded the Coriolis force does not have an impact on the dynamics of a single turbine wake. }
    \label{singleturbu}
\end{figure}

\newpage
\begin{figure}[h]
    \centering
    \includegraphics[width=\textwidth]{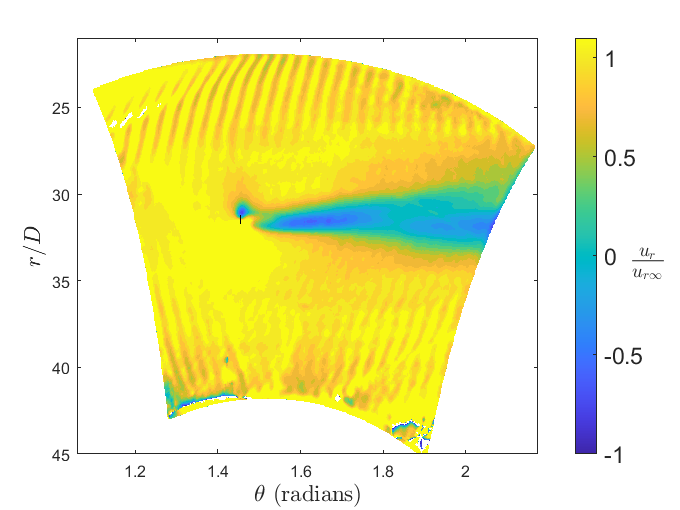}
    \caption{$u_r / u_{r\infty}$ normalized mean radial velocity contour plot in polar coordinates of the single turbine case. The origin is located at the center of the tank. $r$ is normalized by the rotor diameter and $u_r$ is normalized by the free stream radial velocity, $u_{r\infty}$. The magnitude of the normalized radial velocity within the wake of the single turbine is comparable to the magnitude on either side of the wake. This shows that the even though the radial velocity within the wake is reversed, for a single turbine case the magnitude is similar to the surrounding flow and therefore no deflection is seen in the wake. }
    \label{singleturbur}
\end{figure}

\end{document}